\documentclass{aastex7}

\begin{document}

\title{Spectroscopic Observations of Supra-Arcade Downflows}

\author[0000-0001-9726-0738]{Ryan J. French}
\affiliation{Laboratory for Atmospheric and Space Physics, University of Colorado Boulder, Boulder, CO 80303, USA}
\affiliation{National Solar Observatory, 3665 Discovery Drive, Boulder, CO 80303, USA}
\email{ryan.french@lasp.colorado.edu}

\author[0000-0001-8975-7605]{Maria D. Kazachenko}
\affiliation{Laboratory for Atmospheric and Space Physics, University of Colorado Boulder, Boulder, CO 80303, USA}
\affiliation{National Solar Observatory, 3665 Discovery Drive, Boulder, CO 80303, USA}
\affiliation{Department of Astrophysical and Planetary Sciences, University of Colorado Boulder, 2000 Colorado Avenue, Boulder, CO 80305, USA}
\email{maria.kazachenko@lasp.colorado.edu}

\author[0000-0001-8055-0472]{Teodora Mihailescu}
\affiliation{NASA Goddard Space Flight Center, Greenbelt, MD 20771, USA}
\affiliation{Universities Space Research Association, Washington, DC 20024, USA}
\email{tmihailescu@usra.edu}

\author[0000-0002-6903-6832]{Katharine K. Reeves}
\affiliation{Harvard–Smithsonian Center for Astrophysics, 60 Garden Street, Cambridge, MA 02138, USA}
\email{kreeves@cfa.harvard.edu}

\begin{abstract}

Despite their somewhat-frequent appearance in EUV imaging of off-limb flares, the origins of Supra-Arcade Downflows (SADs) remain a mystery. Appearing as dark, tendril-like downflows above growing flare loop arcades, SADs themselves are yet to be tied into the standard model of solar flares. The uncertainty of their origin is, in part, due to a lack of spectral observations, with the last published SAD spectral observations dating back to the \textit{Solar and Heliospheric Observatory} / \textit{Solar Ultraviolet Measurements of Emitted Radiation} (SOHO/SUMER) era in 2003. In this work, we present new observations of SADs within an M-class solar flare on April 2nd, 2022, observed by the Hinode EUV Imaging Spectrometer (EIS) and NASA Solar Dynamics Observatory. We measure Fe XXIV 192.02 \AA\ Doppler downflows and non-thermal velocities in the low-intensity SAD features, exceeding values measured in the surrounding flare fan. The ratio of temperature-sensitive Fe XXIV 255.11 \AA\ and Fe XXIII 263.41 \AA\ lines also allow the measurement of electron temperature, revealing temperatures within the range of the surrounding flare fan. We compare EIS line-of-sight Doppler velocities with plane-of-sky velocities measured by AIA, to construct the 3D velocity profile of four prominent SADs, finding evidence for their divergence above the flare loop arcade -- possibly related to the presence of a high altitude termination shock. Finally, we detect `stealth' SADs, which produce SAD-like Doppler signals, yet with no change in intensity.


\end{abstract}


\accepted{to ApJL, May 2025}

\section{Introduction} \label{sec:intro}

Solar flares are the explosive conversion of magnetic free energy into the acceleration of particles, plasma heating, and light. They are a fundamental driver of space weather at Earth, and it is therefore critical for us to maximize our understanding of these events. Many observed aspects of solar flares can be explained by the standard eruptive model of solar flares, introduced by \citet{Carmichael1964}, \citet{Hirayama1974}, \citet{Sturrock1968} and \citet{Kopp1976}. In this standard model (also referred to as the CHSKP model, after the authors who pioneered it), a rising magnetic flux rope beneath an overlaying magnetic arcade triggers magnetic reconnection along a thin current beneath the flux rope, formed by the inflow of oppositely-orientated magnetic field. This magnetic reconnection is responsible for the conversion of energy we observe as a solar flare. There are, however, a plethora of observed flare phenomena that the standard model cannot yet explain. 

One such example is the existence of Supra-Arcade Downflows, colloquially referred to as SADs. SADs are dark, tendril-like downflows, observed within hot wispy Supra-Arcade fan structures above growing flare arcades \citep{McKenzie1999}. Flare fans, and the SADs within them, are most commonly viewed over or close to the limb, in high-temperature flaring plasma; previously reported in observations from 1) Yohkoh/SXT \citep[`\textit{Soft X-ray Telescope}',][]{Tsuneta1991}, in soft X-rays by e.g. \citet{McKenzie2000}; 2) SOHO/SUMER \citep[`\textit{Solar and Heliospheric Observatory / }Solar UV Measurement of Emitted Radiation',][]{Wilhelm1995}, in far ultraviolet spectra by \citet{Innes2003a,Innes2003b}; and 3) TRACE \citep[`\textit{Transition Region and Coronal Explorer}',][]{Handy1999} in extreme ultraviolet (EUV) imaging by e.g. \citet{McKenzie2009} and \citet{Savage2011}. Since 2010, the Solar Dynamics Observatory's Atmospheric Imaging Assembly \citep[SDO/AIA,][]{Lemen2012} has provided frequent observations of SADs in limb-flares \citep[e.g.][]{Warren2011,Savage2012}, most commonly in the hot 94 \AA\ (Fe XVIII), 131 \AA\ (Fe XXI) and 193 \AA\ (Fe XXIV) EUV channels.
Tendril-like structures, similar in appearance to SADs, have also been observed in astrophysical plasmas, including in supernova remnants, \citep[e.g.][]{Warren2005,Miles2009}

In the solar corona, SADs were initially interpreted as contracting flux tubes, directly connected to plasma outflows at the base of the reconnecting current sheet -- with their low intensity caused by a lower density than the surrounding plasma  \citep[e.g.][]{McKenzie2000,McKenzie2009,Savage2011,Warren2011}. This interpretation was backed up by \citet{Khan2007}, who found that 90\% of the individual SADs analyzed coincided with spikes in X-ray emission -- suggesting that the SADs are linked to the energy release process. \citet{Samanta2021} found a similar correlation between the incidence of SADs and spikes in EUV emission. Similar suspected reconnection outflows have been observed within solar flare plasma sheets (heated turbulent plasma around likely current sheet sites), such as those presented by \citet{Longcope2018}.
In this scenario, SADs would therefore be a product of non-continuous bursty reconnection in the flare current sheet, similar to that eluded to by observations of quasi-periodic pulsations \citep[see reviews by][]{nakariakov2009, vandoorsselare2016, Kupriyanova2020, Zimovets2021}. This understanding of SADs was reinterpreted in \citet{Savage2012}, who utilized the higher resolution (than-previously-available) AIA imaging to reveal that the dark SAD features are preceded by narrow contracting loops (named Supra-Arcade Downflowing Loops, or SADLs). Following these observations, \citet{Savage2012} suggested that the contracting SADLs were the reconnection outflows, and dark SADs simply the low-density wake left behind the contracting loops. This new scenario was backed up by modeling work in \citet{Scott2016}.

SADs typically travel at speeds of $\approx100-300$ km/s, with observed velocities of 100-200 km/s by \citet{McKenzie1999}, 30-260 km/s by \citet{McKenzie2009}, 144 km/s (initial velocity) by \citet{Warren2011}, and 50-200 km/s by \citet{Xie2022}. These speeds are well below the expected Alfv\'en speeds in the lower solar corona (around 1000 km/s), and therefore sub-Alfv\'enic in nature. This presents a contradiction to the idea that SADs are connected to outflows from the reconnecting current sheet, as these flows are predicted to occur close to the local Alfv\'en speed. Some simulation works \citep[e.g.][]{Shen2011} find that fine small-scale turbulent structures can propagate from the reconnection site slower than the bulk Alfv\'enic reconnection outflows, but not slow enough (relative to the Alfv\'en speed) to explain the bulk of observed SAD speeds. This discrepancy was addressed in MHD simulation work from \citet{Shen2022}, who suggest a high-altitude termination shock, located above the flare arcade and below the primarily reconnection site, may rapidly decelerate the Alfv\'enic reconnection outflows to speeds closer to those observed in SADs. Observational evidence for such termination shock structures exist for only a handful of case studies \citep[e.g.][]{Chen2015,Chen2019,Polito2018,French2024a}.

Part of the uncertainty in the nature in SADs is due to the lack of spectroscopic observations of the structures.
Although their appearance in full-disc EUV imagers are somewhat numerous, spectral observations of SADs are very rare, in part due to the difficulty of catching them with small field-of-view (FOV) instruments. There are a limited number of EUV spectroscopic datasets of off-limb flare fans \citep[e.g.][]{Reeves2020,French2024b} and plasma sheets \citep[e.g.][]{Warren2018,Longcope2018,French2020,Shen2023} from recent years, but spectral measurements of individual SAD downflow structures have not been published since the SUMER era \citep{Innes2003a,Innes2003b}. As a result of this lack of spectroscopic measurements, the exact nature of SADs remains uncertain. In this work, we present the first spectra of SADs taken by the Hinode EUV Imaging Spectrometer \citep[EIS, ][]{Culhane2007}, and the first spectroscopic measurements of individual SADs anywhere since the SUMER era. We analyze the Fe XXIV 192.02 \AA\ intensity, Doppler velocity and non-thermal velocities of the SAD structures, and measure their electron temperature using the Fe XXIV 255.11 \AA\ / Fe XXIII 263.41 \AA\ line pair. Finally, we compare the EIS line-of-sight (LOS) velocity measurement with plane-of-sky (POS) measurements from AIA.

\section{Observations} \label{sec:Observations}

\begin{figure*}
\centering
\includegraphics[width=18cm]{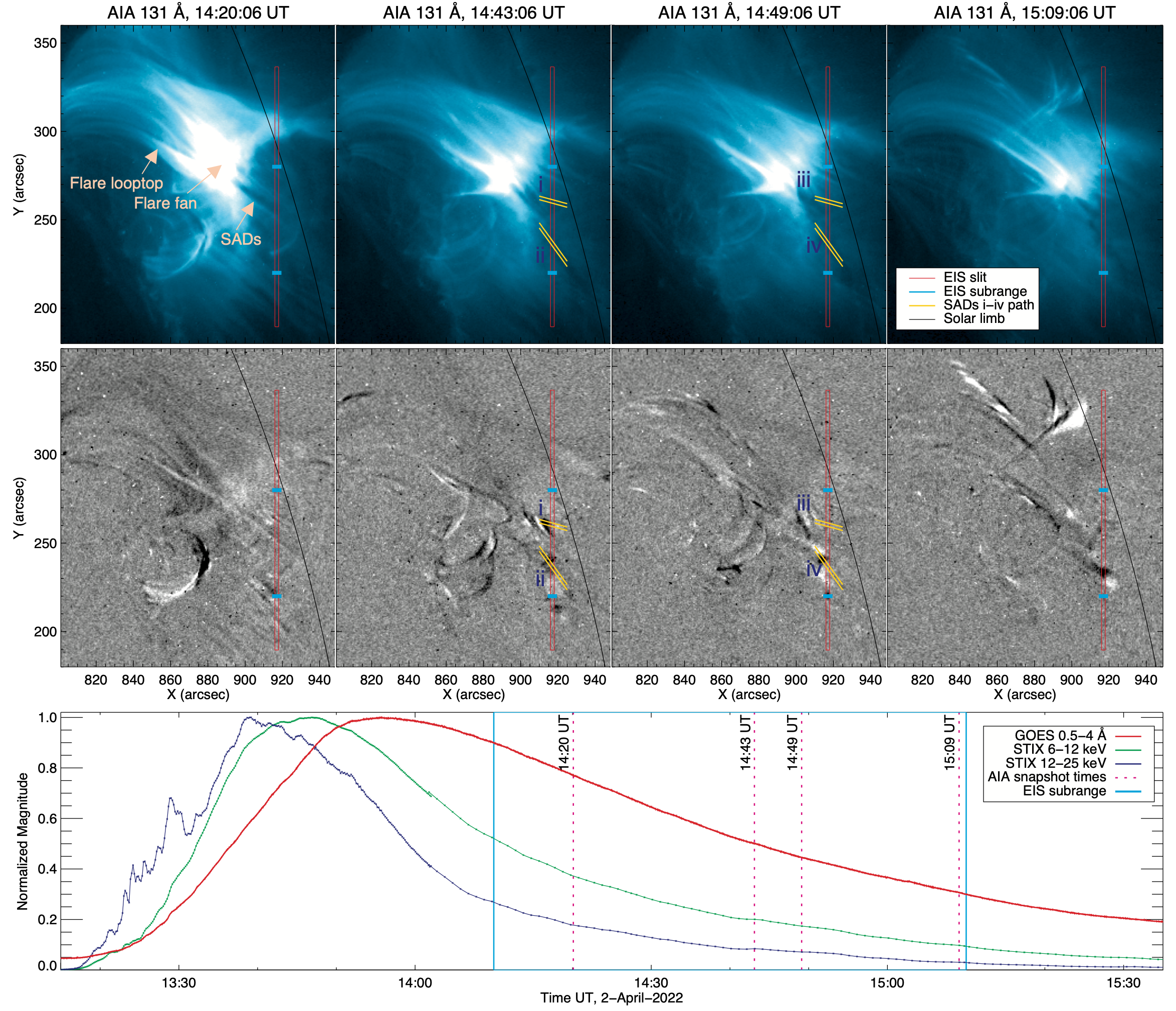}
\caption{Top row: AIA 131 \AA\ snapshots of the flare evolution, at times where individual Supra-Arcade Downflow (SAD) structures cross the EIS slit. The red slit identifies the location of the sit-and-stare EIS slit, and small blue markers the subrange of the EIS raster analyzed in Figures \ref{fig:eis_all} and \ref{fig:AIA_slits}. The black line marks the location of the solar limb. The yellow cross-sections mark the path of SADs i-iv, four individual SADs we analyze in this study.
Middle row: Same as top row, instead showing AIA 131 \AA\ difference images with a one minute running base.
Bottom row: Light curves for GOES 0.5-4 \AA, STIX 6-12 keV and STIX 12-25 keV emission. Vertical cyan lines mark the start and end of the time period} analyzed in this work. Vertical dashed red lines show the time of the AIA snapshots presented in the upper rows. The movie version of this Figure shows the continuous evolution of the same parameters from 14:00-15:35 UT.

\label{fig:AIA131}
\end{figure*}

Hinode EIS is a spectrometer observing EUV emission lines from cool active region loops (e.g. Fe VIII) to hot flare emission (e.g. Fe XXIV) temperatures. On April 2nd 2022, EIS observed an M3.9-class solar flare towards the northwestern limb, with a 2\arcsec-wide north-south oriented slit, in a sit-and-stare observation sequence of 16 s cadence and 1\arcsec\ pixel size (along the slit). Figure \ref{fig:AIA131} highlights the location of the EIS sit-and-stare slit relative to the flare and solar limb, overplot on AIA 131 \AA\ snapshots of the flare evolution. 
EIS captured the full duration of the flare, including a filament eruption \citep{Janvier2023}, late flare loop growth, and a hot supra-arcade fan.

The AIA images in Figure \ref{fig:AIA131} show a flare loop arcade, with a bright flare fan above it. This flare fan is visible both off-limb (the solid black line), and on disk, against the darker background quiet Sun. Dark streaks are visible within the flare fan, stretching from the center-right of the brightest part of each image, towards the lower-right (marked in the Figure). Some of these darker structures are static, but some are dynamic -- the latter of which are SADs. The second row of Figure \ref{fig:AIA131} present difference imaging (over a 1 minute running base) of the panels above. In these panels, the dynamic SADs appear black, representing the decrease in intensity created by the downflowing structures. At several times, and in multiple locations, dark SADs (seen clearly in the difference imaging panels) cross the location of the EIS slit. The paths of the clearest of these are marked by the yellow-lines; cross-sections along the direction of SAD flow, which we analyze later in this work. At each fixed time frame, there is a SAD present somewhere within the yellow SAD-path cross-section.

The bottom row of Figure \ref{fig:AIA131} shows the light-curve of the flare, in the 0.5-4 \AA\ channel of the Geostationary Operational Environmental
Satellite (GOES) X-ray sensor (XRS), and 6-12 and 12-25 keV channels of the Solar Orbiter \citep{Muller2020} Spectrometer Telescope for Imaging X-rays \citep[STIX,][]{Krucker2020}. Solar Orbiter was located $\approx110^{\circ}$ west of Earth during the observation period, so observed the flare on disk \citep[Figure 1 of][show a detailed plot of Solar Orbiter's location during this event]{Janvier2023}.
Comparing the AIA snapshots of strong SADs with the timing of the flare, we find the SADs are strongest beyond the peak of the event. This is fairly typical, with SADs often observed during the decay phase of flares \citep{Warren2011}.

\section{Spectral Analysis} \label{sec:spectral_analysis}

\begin{figure*}
\centering
\includegraphics[width=15.5cm]{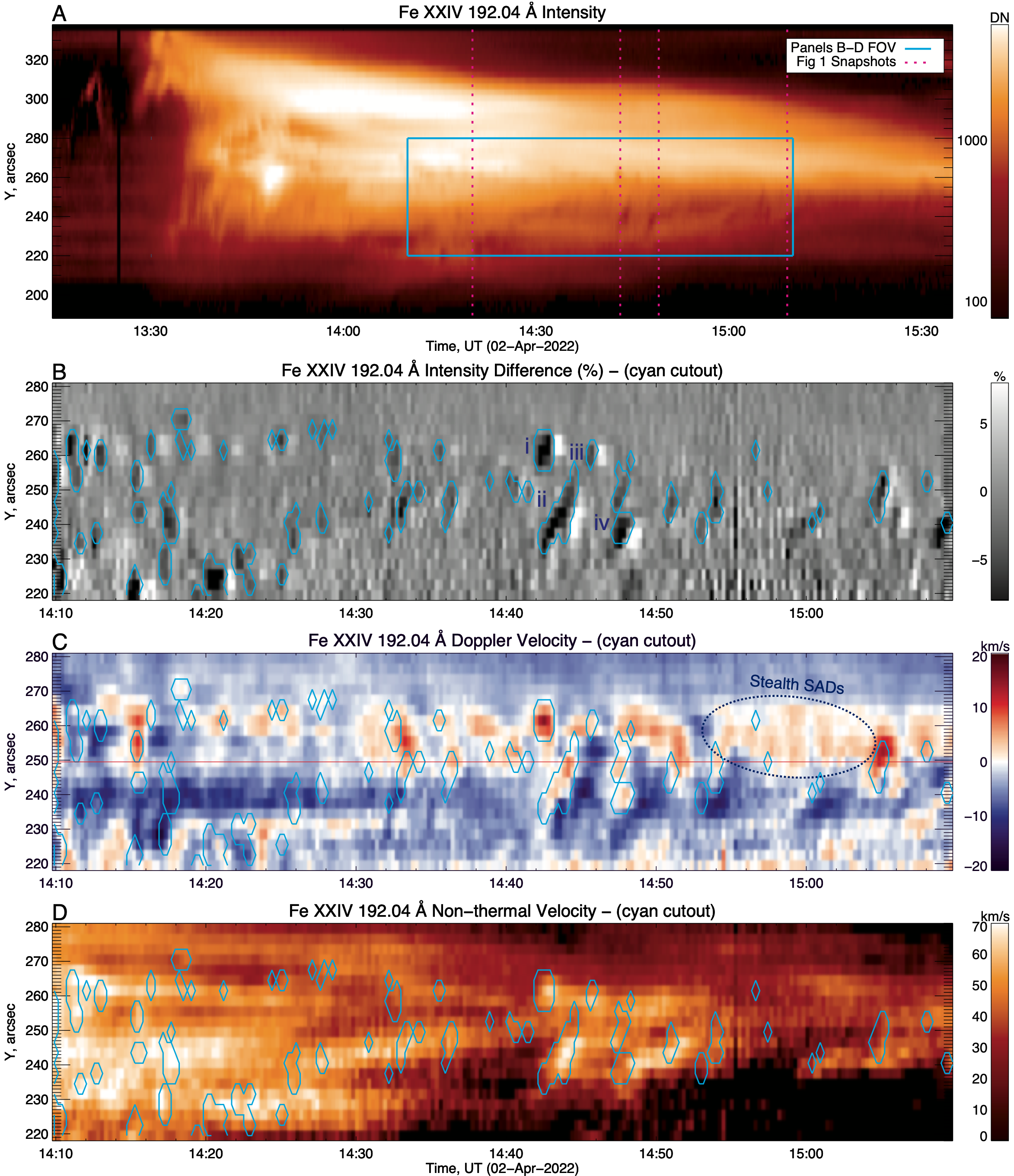}
\caption{
A: EIS Fe XXIV 192.02 \AA\ intensity sit-and-stare maps for full time range plotted in Figure \ref{fig:AIA131} (bottom), along the full EIS slit. The cyan box marks the sub-range FOV presented in subsequent panels. The vertical red dashed lines mark the location of AIA snapshots in Figure \ref{fig:AIA131}.
B-D: EIS Fe XXIV 192.02 \AA\ intensity running percentage difference, Doppler velocity, and non-thermal velocity, respectively. The cyan contours mark the location of our individual SADs identified from the running difference map (B). The four primary SADs analyzed in further detail are labeled i-iv in panel B. The horizontal red line in panel C marks the boundary between north/south SADs explored in Figure \ref{fig:spectra}. The location of example `stealth SADs' is plotted in panel C.
}
\label{fig:eis_all}
\end{figure*} 

\begin{figure*}
\centering
\includegraphics[width=18cm]{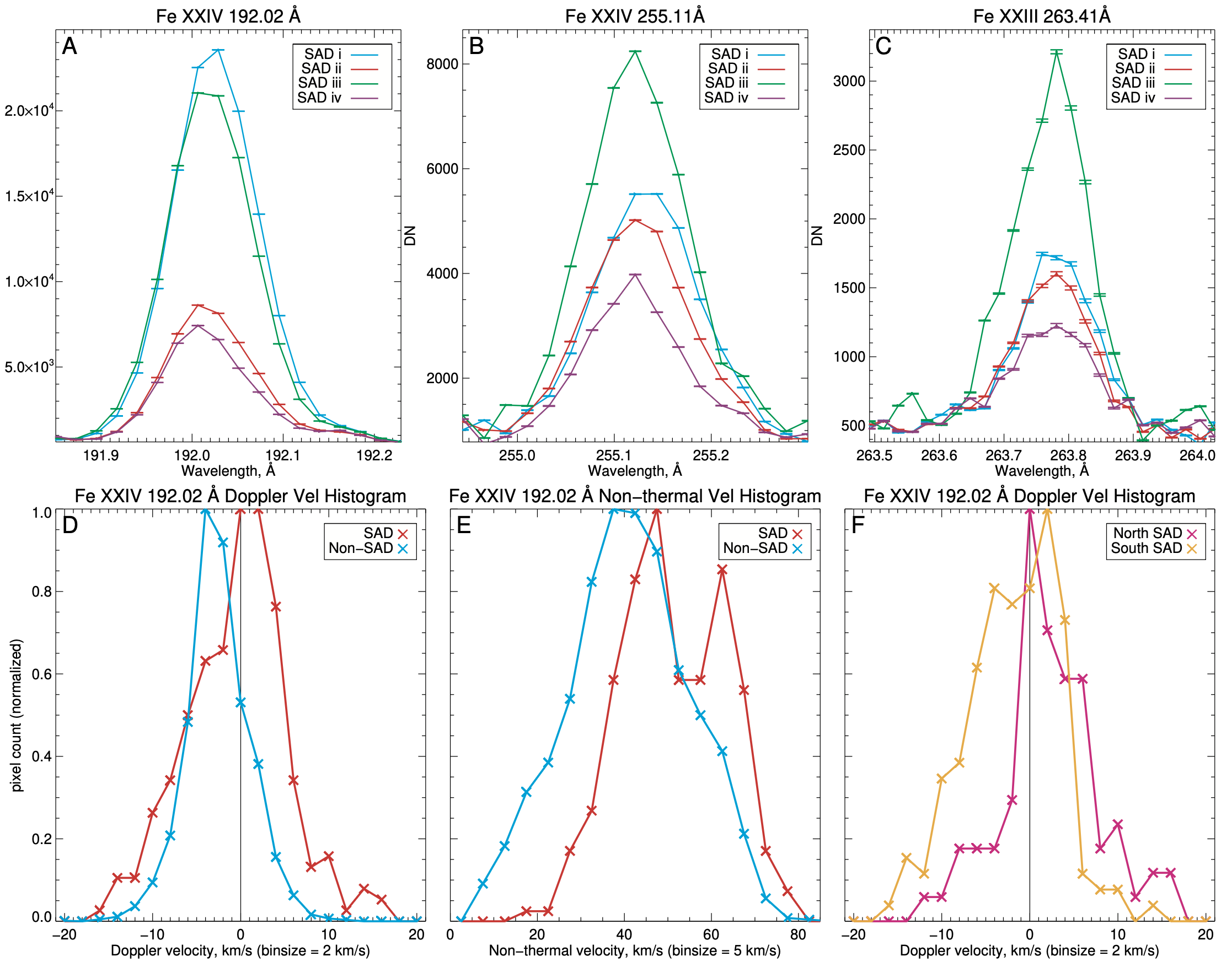}
\caption{
A-C: Spectroscopic line profiles for key SADs i-iv, in Fe XXIV 192.02 \AA\ (left), Fe XXIV 255.11 \AA\ (center) and Fe XXIII 263.41 \AA\ (right).
D-E: Histograms for Fe XXIV 192.02 \AA\ Doppler velocity and non-thermal velocity, comparing values between all SAD and non-SAD pixels, as per contours in Figure \ref{fig:eis_all}. F: Comparison of Doppler velocities between northern and southern SADs, defined by their position above/below the orange horizontal line in Figure \ref{fig:eis_all}.
}
\label{fig:spectra}
\end{figure*}  

\subsection{Fe XXIV Intensity}

Figure \ref{fig:eis_all}A shows the stitched sit-and-stare raster (of three individual, but immediately-subsequent observations, beginning at 13:05, 14:06 and 15:06 UT respectively) for Fe XXIV 192.02 \AA\ intensity. Hot, warm and bad pixels are removed using the well-established solarsoft $eis\_prep$ routine, (and ignored during the binning processes). The Fe XXIV 192.02 \AA\ line is blended with an Fe XI line \citep{Young2007}. However, the \textit{CHIANTI Spectral Synthesis Package \citep{Dere1997,Dufresne2024}} reveals that for a typical flare differential emission measure, the Fe XXIV line is three orders of magnitude stronger than the Fe XI line -- and thus a negligible source of uncertainty.

We calculate the line intensity by fitting a single Gaussian curve to the measured spectra. We find a single Gaussian fit works well in this dataset, with no pixels of interest deviating noticeably from a Gaussian shape. To improve the signal-to-noise of the data, we apply a 1x3 pixel binning, reducing the spatial resolution along the slit to 3\arcsec, but maintaining the original 16 second cadence. The start of the raster catches the filament eruption, but the prolonged bright Fe XXIV emission originates from the bright, hot flare fan structure visible in AIA 131 \AA\ in Figure \ref{fig:AIA131}A. The four AIA snapshots in Figure \ref{fig:AIA131} were chosen for times at which SADs visibly cross the location of the EIS slit. The times of these snapshots are marked by vertical dashed lines in the Figure \ref{fig:eis_all}A EIS raster. At these marked times, we see short-lived dark structures in the Fe XXIV EIS raster, at locations along the slit matching their occurrence in AIA images. These dark features, of which there are many visible within the cyan box in Figure \ref{fig:eis_all}A, are produced by SADs crossing the one-dimensional EIS slit. There is a drop in intensity as they cross the slit, before intensity levels return to that of the brighter flare fan after the SADs pass. Subsequent Figure \ref{fig:eis_all} panels B-D present maps of further Fe XXIV parameters, cropped in time and along the slit to match the field-of-view of the cyan box in panel A.

Figure \ref{fig:eis_all}B presents the Fe XXIV intensity difference as a percentage, calculated by subtracting a pixel intensity from the pixel immediately before it, and dividing by the original value. In this panel, the most prominent SADs appear as a pair of horizontally-spaced negative-to-positive (black-to-white) difference features. The negative part of the feature (appearing black) is the SAD, as it crosses the slit and lowers the observed intensity. This intensity decrease is then followed by a subsequent increase (appearing white), after the SAD has passed the slit completely and intensities return to background levels. This black-white pairing is crucial in identifying the SADs, to distinguish between the short-term intensity drop from the transient SADs, and the longer-lasting intensity drops possible from other dimming processes. The intensity difference map in Figure \ref{fig:eis_all}B reveals many such black-white pairs -- some prominent,  some much fainter, with many difficult to see in the (non difference-imaged) intensity map in panel A. We automate the detection of these SADs by identifying regions of three subsequent timesteps of a dimming below -1.8\%, with a minimum dimming in that region below 3\%, immediately followed by a brightening timestep, for each pixel along the slit. We exclude detections within the high noise region around 220-230\arcsec\ from 14:50 UT onward. All of the automatically-detected SADs are marked as cyan contours in panels B-D of Figure \ref{fig:eis_all}. A cluster of four prominent high-contrast SADs occurring close together in space and time, are labeled as SADs i-iv. These SADs are also seen in the AIA snapshots of Figure \ref{fig:AIA131}.  Figure \ref{fig:spectra}A shows the Fe XXIV 192.02 \AA\ spectra of these four SADs (pixels within the contour are binned to a single spectra). We also take a deeper look into the spectral (sections \ref{sec:Doppler}-\ref{sec:Temperature}) and plane-of-sky (Figure \ref{fig:AIA_slits}) properties of these individual SADs, later outlined in Table \ref{table:table}.

\subsection{Fe XXIV Doppler Velocity}\label{sec:Doppler}

Figure \ref{fig:eis_all}C presents a map of Fe XXIV 192.04 \AA\ Doppler velocity. Doppler velocity is calculated by measuring the red/blue shift of the fitted Gaussian centroid from the Fe XXIV rest wavelength. We determine the rest wavelength of Fe XXIV 192.04 \AA\ from calibration to quiet Sun emission of the nearby Fe XII 192.39 \AA\ line. The point-to-point uncertainties of the Doppler velocities are in the $1-1.5$ km/s range, allowing us to safely detect \textit{relative} velocity changes above this level. There is, however, a larger $\approx5$ km/s uncertainty in the rest wavelength of EIS measurements, which is a blanket uncertainty applied uniformly to the dataset (but does not affect velocity variations throughout the data. In-depth discussions on the wavelength uncertainties of EIS data can be found in \citet{Mariska2010} and \citet{Young2012}. 

The majority of the Doppler map is blue shifted, indicative of hot Fe XXIV plasma rising towards us. This is expected, as the hot flare fan expands above the new flare loops forming at progressively higher altitudes. Within the bulk blue-shifted plasma, we see regions of concentrated red shift. In the top half of the map, nearly all the SAD contours coincide with regions of red-shifted plasma, which is especially true for the higher-contrast SADs. This is indicative that the dark SADs themselves are red-shifted, and not the plasma above or beneath them. Figure \ref{fig:spectra}D presents a histogram comparing the Doppler velocity of SAD versus non-SAD pixels in the submap FOV, revealing more and stronger red-shifted pixels within the SADs, than in the non-SAD fan.

Examining the Doppler map, we see fewer red-shifted SADs in the lower portion of the map, than the upper portion, with some SAD pixels even exhibiting blue shift. The histogram in Figure \ref{fig:spectra}F compares the Doppler velocities of SAD pixels above and below the horizontal red line in Figure \ref{fig:eis_all} Doppler map. The histogram reveals the trend of stronger and more numerous blue-shifted SADs in the south of the FOV, than in the north. At first, this result seems unexpected, as SADs are downflowing towards the top of the flare arcade. However, this velocity discrepancy can be explained by the viewing angle and geometry of the flare, where the spatial difference in red/blue shift is likely a result of SADs diverging above the flare loop arcade. 
Curiously, in addition to the red-shift features aligning with the SAD contours, we also see many other red-shift structures. These structures are similar in appearance to those belonging to the visible SADs, but relate to no corresponding drop in intensity. Strong examples of these features are seen in Figure \ref{fig:eis_all}C from 14:55-15:03 UT at $Y=250-265$\arcsec. This reveals that not all SAD-like flows produce an observable drop in intensity, yet do still produce a clear red-shift signal above the point-to-point Doppler uncertainty. To our knowledge, the existence of these `stealth SADs' have not previously been predicted or observed.

Doppler velocities measure the bulk flow of plasma. In situations with two or more individual bulk flows within the LOS, a single Gaussian curve may not suitably fit the spectral profile. The spectra may have an enhanced blue or red wing, or even be comprised of two distinct Gaussian shapes. In the spectral analysis of SADs from \citep{Innes2003b}, the authors find a strong blue-shifted wing in the Fe XXI SAD spectra, much faster than velocities seen in the brighter surrounding plasma. They interpret this enhanced wing as strong downflows of 800-1000 km/s towards the loop arcade (the flare was over the limb, so downflows were observed as blue-shift, not red-shift). In our Doppler analysis of the April 2nd 2022 event we present in this paper, we find no clear evidence of similar high-velocity wing enhancements. Figure \ref{fig:spectra} shows the binned Fe XXIV 192.02 \AA\ spectra for named SADs i-v, which show no clear features in the wing. Additionally, the manual inspection of much of the SAD pixel spectra reveals that the profiles are suitably represented by a Gaussian fit.

\subsection{Fe XXIV Non-thermal Velocity}

Non-thermal velocities are the velocity component of excess broadening in an emission line, beyond the thermal and instrumental broadening. It is often considered to be a result of unresolved plasma flows (with no preferential direction) in the LOS, such as that produced by waves or turbulence \citep{Doschek2014}.
Figure \ref{fig:spectra}D shows a map of non-thermal velocity for our region of interest. We see enhanced non-thermal velocities around the SAD contours -- with some values reaching 70 km/s and higher. The spatial alignment between the SAD contours and regions of enhanced non-thermal velocity is not perfect, but the correlation is verified by the histogram plotted in Figure \ref{fig:spectra}E. In this panel, we compare the non-thermal velocity of SAD to non-SAD pixels. We find SAD pixels have a minimum non-thermal velocity of 20 km/s, whereas the distribution of non-SAD pixels falls to zero. As well as fewer low-velocity pixels, we also find more pixels with stronger non-thermal velocities. Stronger non-thermal velocities within the SADs suggest a presence of enhanced waves or turbulence within the structures. For the highest contrast SADs, including the labeled SADs i-iv, the non-thermal velocity map (Figure \ref{fig:spectra}D) shows that non-thermal velocity enhancements trail in time behind the intensity drops (cyan contour), even persisting at the EIS slit location after the SAD has passed. This time lag (and persistence) of non-thermal velocity could be an indication that the SADs are causing turbulence as they move through the flare fan.

\subsection{Temperature}\label{sec:Temperature}

The spectra of the SADs in the Fe XXIV 255.11 \AA\ and Fe XXIII 263.41 \AA\ lines allow us to measure the electron temperature of the SADs. Fe XXIV 192.02 \AA\ is not the only wavelength in which the SADs appear in our EIS observations. Although the signal of them is very weak, the SADs are also visible in Fe XXIV 255.11 \AA\ and Fe XXIII 263.41 \AA. Strictly speaking, in the un-binned raster at these wavelengths, the SADs are only visible as dark pixels with a \textit{lack} of signal, nestled within the bright high-signal flare fan. Unlike the flare fan (with a strong signal-to-noise), the spectra within the SAD pixels are too noisy to reveal any profile. To combat this, we rebin the EIS raster to obtain an average spectra across each of the four clear example SADs labeled in Figure \ref{fig:eis_all} and Figure \ref{fig:AIA_slits}, based on the position of the SADs in the clearer Fe XXIV 192.02 \AA\ maps (adjusted for the y-offset between EIS detectors). This process successfully increases the signal-to-noise, revealing clear Gaussian profiles for the Fe XXIV 255.11 \AA\ and Fe XXIII 263.41 \AA\ lines in our four labeled SADs. These line profiles are presented in panels B and C of Figure \ref{fig:spectra}. Because adequate signal was only reached with significant binning, we are unable to produce a map of the spectral properties at these wavelengths, (similar to that shown for Fe XXIV 192.02 \AA\ in Figure \ref{fig:eis_all}). 

From their intensities alone, the Fe XXIV 255.11 \AA\ and Fe XXIII 263.41 \AA\ lines are highly valuable, due to the fact that they make up a temperature-sensitive pair. By comparing the two line intensities, we cross-reference the ratio of the two lines to the theoretical Fe XXIV 255.11/Fe XXIII 263.41 \AA\ ratios provided by the CHIANTI database \citep{Dere1997,Dufresne2024}. Following this comparison, we provide the first ever recorded electron temperature measurements of SADs from EUV line ratios. The temperature values for the four SADs, and background non-SAD fan (selected by a tight square encompassing SADs i-iv), are listed in Table \ref{table:table}. For SADs i-iv, we measure temperature ranges of 12.2 - 13.4 MK, compared to a nearby non-SAD fan temperature of 12.6 MK.  

In previous studies, temperature estimates of SADs have been obtained from differential emission measure (DEM) analysis, with large variations between different works. Using DEM analysis of AIA data, \citet{Li2021} find SAD temperature ranges of 7-10 MK, and \citet{Brose2022} a range of 6.5-11 MK, both $10-20\%$ below our measurements from EUV line ratios. Using analysis of both AIA and hotter XRT observations, \citet{Hanneman2014} impose an upper SAD temperature limit of 22 MK, significantly lower than 84 MK and 22-29 MK values obtained in modeling efforts from \citet{Maglione2011} and \citet{Cecere2012} respectively. This $<$22 MK limit is compatible with our measurements in the 12.2-13.3 MK range. \citet{Hanneman2014} also find SAD temperatures close to that of the surrounding flare fan, a result also seen in our analysis. \citet{Xie2023} find SAD temperatures of 6-8 MK, slightly cooler than that of the surrounding fan.
In another study of AIA analysis, \citet{Tan2022} find a broad SAD temperature range of $10^6-10^{7.5}$ (1-31.6) MK, a broad range also in line with our measurements. 
We note that our line ratio measurements utilize binning over each individual SAD structure. Because the temperature of SADs are known to vary across an individual SAD area, \citep[with a maximum value in the front region, e.g.][]{Savage2012,Hanneman2014}, and also increase with time as the SAD propagates \citep{Reeves2017,Xue2020}, we are therefore measuring the average temperature of the SADs in both space and time as it crosses the EIS sit-and-stare slit.

The measurement of electron temperature through EUV line ratios makes the assumption of ionization equilibrium. Work by \citet{Kawate2016} investigates whether such an assumption is valid for EIS observations of solar flares, finding that only $\sim1.4\%$ of pixels fail to meet this assumption. They therefore conclude that, given the timescale of EIS exposures, the ionization equilibrium assumption is valid in most cases. We make this same assumption in our work, whilst acknowledging that departures from ionization equilibrium could still introduce some unquantified uncertainties into these electron temperature measurements.

\section{Plane-of-Sky Analysis} \label{sec:POS_analysis}

\begin{figure*}
\centering
\includegraphics[width=15.5cm]{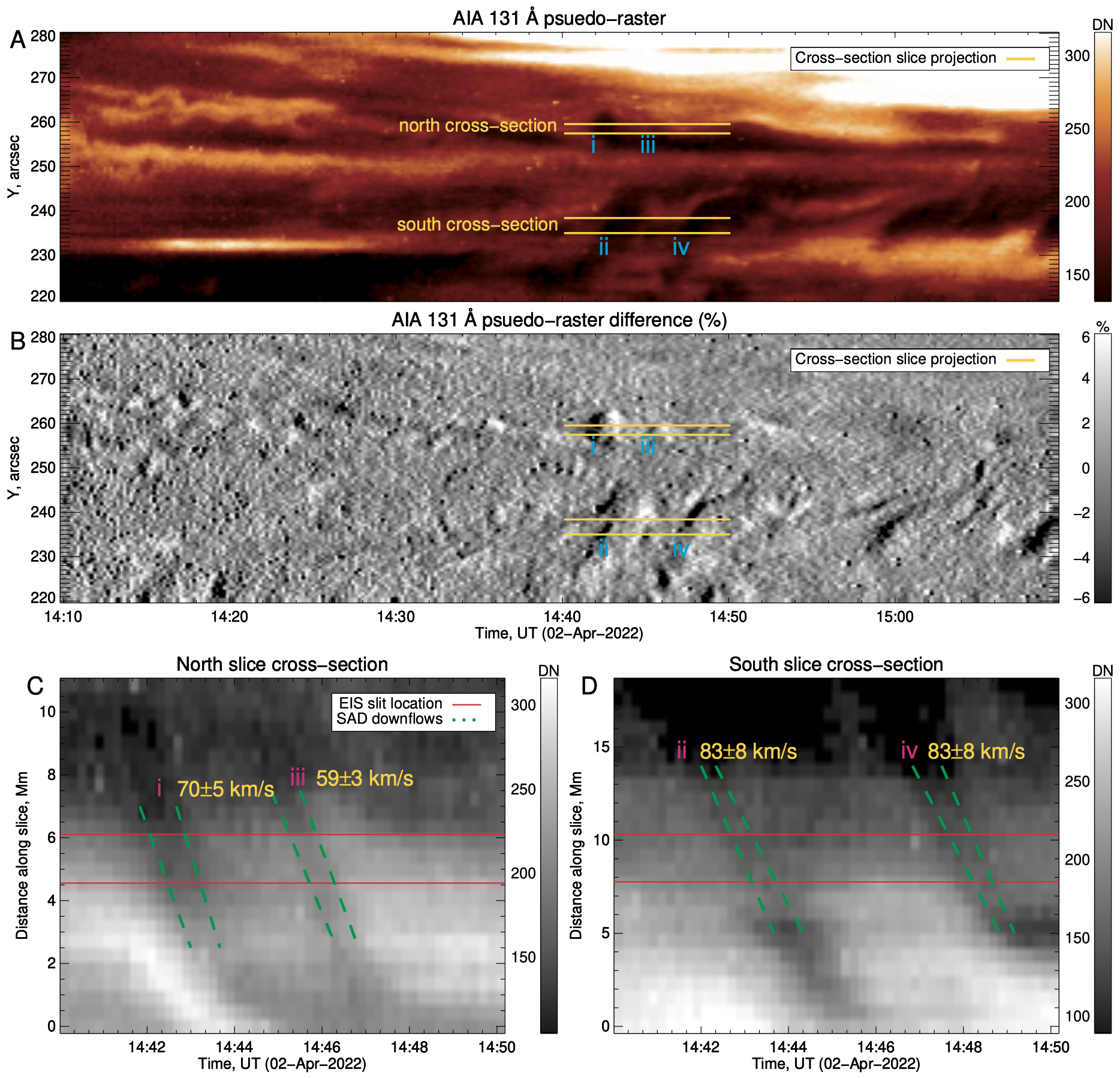}
\caption{
A: AIA 131 \AA\ emission along the EIS slit subrange, creating a pseudo-raster. Our key SADs of interests are labeled i-iv. Yellow horizontal lines show the \textit{projection} across the slit position of the north and south SAD path cross-section slices plotted in Figure \ref{fig:AIA131}, for the duration of the cross-sections plotted in the bottom panels of this current Figure.
B: Difference imaging (as a percentage) of panel A, subtracting each column from the one prior, and dividing by the original value, at the full AIA cadence of 12 seconds.
C/D: Stack plots showing AIA 131 \AA\ emission along the north (C) and south (D) cross-sections slices presented in Figure \ref{fig:AIA131}. We capture the downflows of key SADs i-iv, and track their velocity across the EIS slit position (red lines).
}
\label{fig:AIA_slits}
\end{figure*}

\begin{table}
    \centering
    \begin{tabular}{ccccccc}
       $ SAD \#$  & $V_{los}$ & $V_{pos}$ & $V_{tot}$ & $Temp$ & $V_{NT}$ \\
\textbf{i} & $8.7\pm1.0$ km/s & $70\pm5$ km/s & $70.1\pm5.4$ km/s &  $13.4\pm0.2$ MK & $49.2\pm2.6$ km/s \\
\textbf{ii} & $5.0\pm0.9$ km/s & $83\pm8$ km/s & $82.6\pm7.5$ km/s &  $13.3\pm0.2$ MK & $59.4\pm2.3$ km/s \\
\textbf{iii} & $2.1\pm0.3$ km/s & $59\pm3$ km/s & $59.0\pm3.4$ km/s &  $12.2\pm0.2$ MK & $52.7\pm5.3$ km/s \\
\textbf{iv} & $3.1\pm0.6$ km/s & $83\pm8$ km/s & $82.5\pm7.5$ km/s &  $12.7\pm0.2$ MK & $46.2\pm2.4$ km/s \\
\textbf{Non-SAD} & $-2.3\pm0.2$ km/s & $N/A$ & $N/A$ & $12.6\pm0.2$ MK & $42.1\pm0.6$ km/s \\

    \end{tabular}
    \caption{Line-of-sight and plane-of-sky parameters for labeled SADs i-iv and a sample non-SAD region. Specific column parameters are: Fe XXIV 192.02 \AA\ line-of-sight Doppler velocity ($V_{los}$), plane-of-sky velocity ($V_{pos}$), total velocity ($V_{tot}=\sqrt{V_{los}^2+V_{pos}^2}$), line-of-sight angle ($\theta_{LOS} = \tan(V_{los}/V_{pos})$), to the nearest degree, Fe XXIII 263.41 \AA\ / Fe XXIV 255.11 \AA\ temperature, and Fe XXIV 192.02 \AA\ non-thermal velocity for labeled SADs i-iv in previous figures. 
    }
    \label{table:table}
\end{table}

In addition to the LOS velocity values determined by the Doppler analysis of Fe XXIV 192.02 \AA\ lines, we can also measure the POS velocities of SADs from AIA imaging (not possible from the sit-and-stare EIS raster). Figure \ref{fig:AIA_slits}A shows a pseudo-raster of AIA 131 \AA\ observations, showing the evolution of AIA 131 \AA\ emission along the location of the EIS slit (between the vertical red lines in Figure \ref{fig:AIA131}) with time, for the same time duration used for the EIS raster sub-range plots in Figure \ref{fig:eis_all}. The 131 \AA\ pseudo-raster bares some resemblance to the slightly hotter EIS Fe XXIV 192.04 \AA\ intensity raster ($\log(T)=7$ versus $\log(T)=7.3$), including the presence of black streaks created by dark SADs crossing the slit location. The SADs are better seen in Figure \ref{fig:AIA_slits}B, which presents a percentage running difference version of panel A. In this panel, each column is subtracted from the previous, and divided by the original, at the full AIA cadence of 12 seconds. This bears similarity to Figure \ref{fig:eis_all}, which presents the same analysis for EIS data. The four labeled SADs (i-iv) are visible here, alongside other fainter features. The smallest labeled SAD, SAD iii, is less obvious in the AIA data (compared with its appearance in the EIS data), but is still present.

To determine the POS velocities of the SADs, we must measure their speed along their direction of travel.
The yellow slices in the top and middle panels of Figure \ref{fig:AIA131} mark the location and direction of our four labeled SADs i-iv. The location of these yellow slices are also projected onto the AIA 131 \AA\ pseudo-raster in Figure \ref{fig:AIA_slits}. Their projection on the pseudo-raster is not the full extent of the cross-section, only the intersection between the cross-section and location of the EIS slit.
For each of these two slices, we plot a time-distance diagram to capture the full component of their POS velocity. These time-distance plots along the direction of travel are presented in the panels C and D of Figure \ref{fig:AIA_slits}. The horizontal red lines mark the position of the EIS slit along the cross-section slice. Note -- the y-axis of each time-distance plot has a different axis range (with the width of the projected EIS slit location differing between the two), due to the difference in angle of the slices.
In the height-time diagrams, we see the SADs i-iv moving as dark features flowing along our slice, crossing through the location of the EIS slit. We manually fit linear fits to the downflows, to measure the POS velocity of the SADs. To provide a measurement of the uncertainty, we fit two lines for each SAD -- one on either edge of the dark downflow feature, to provide an uncertainty range for the POS velocity values. These POS velocities for SADs i-iv are located in the same point in space and time as the Doppler measurements of the same SADs (Figure \ref{fig:eis_all}). By comparing the POS and LOS (Doppler) velocities, we can calculate the total magnitude and direction of the velocity for each SAD. These values are populated in Table \ref{table:table}. For the LOS velocity measurements, we take the mean and standard error of Doppler velocities within each SAD contour.
For our four SADs of interest, we find total velocities of 59-83 km/s, orientated close to the plane of sky. (Because of the systematic $\approx5$ km/s uncertainty applied uniformly to the EIS raster, we cannot estimate the specific POS angle with any accuracy). These values are similar to, although on the lower end of, the POS downflows found from previous imaging of SADs \citep{McKenzie1999,McKenzie2009,Xie2022}. In Figure \ref{fig:spectra}F, we find a pattern that SADs in the north are more likely to exhibit stronger red shifts than those in the south, suggesting that the structures are diverging above the flare looptop. 

\section{Discussion} \label{sec:concl}

\begin{figure*}
\centering
\includegraphics[width=10.5cm]{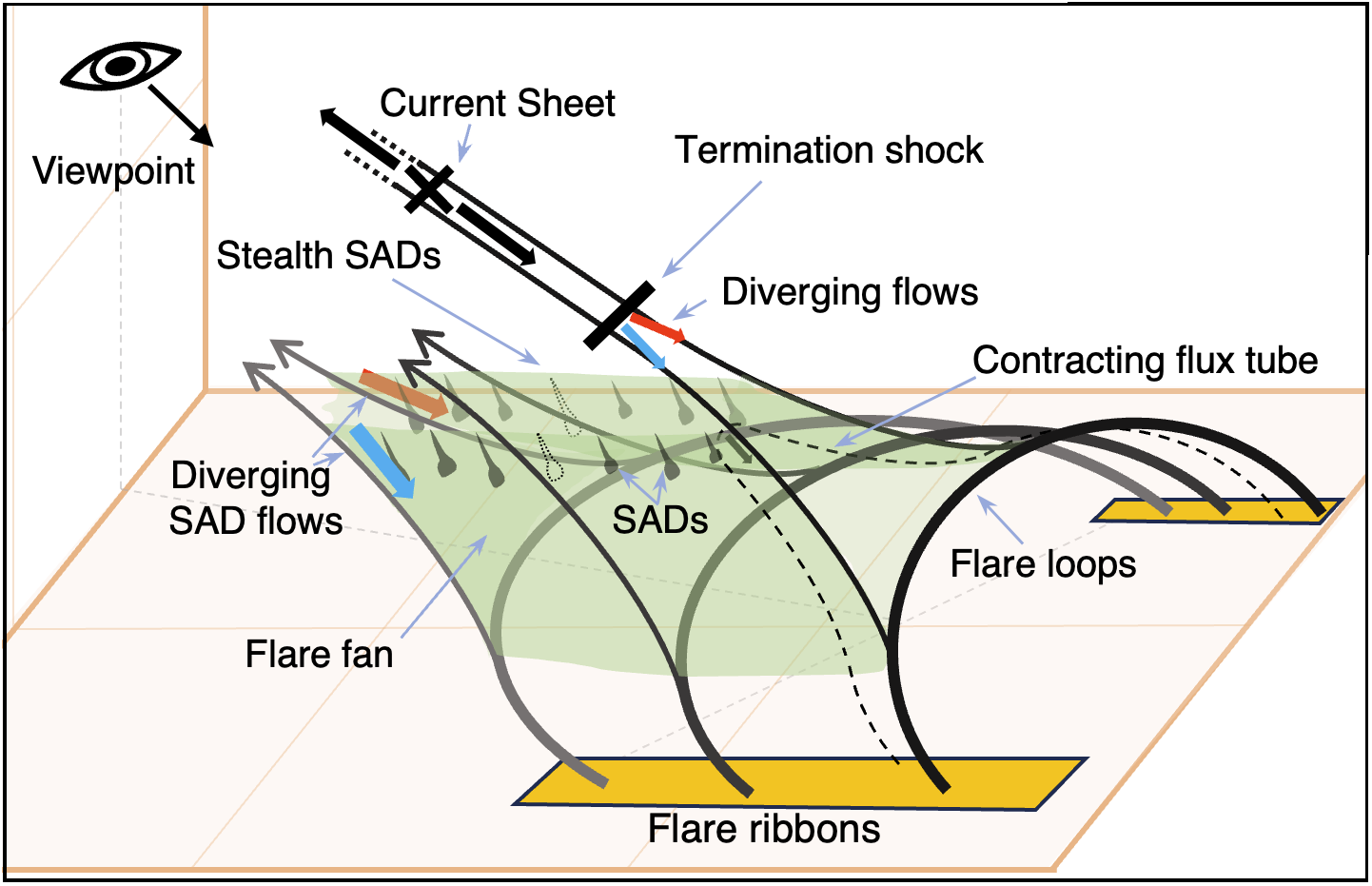}
\caption{
A cartoon depicting the simplified geometry of the of the April 2nd 2022 M-class solar flare and SADs, relative to the suspected location of the higher altitude reconnection site and possible termination shock. The viewing geometry explains the observed distinction in Doppler velocities between the northern and southern SADs.
}
\label{fig:SAD_cartoon}
\end{figure*}  

In this work, we have presented the first spectroscopic observations of SADs since the SUMER era \citep{Innes2003a,Innes2003b}. The Fe XXIV 192.02 \AA\ sit-and-stare measurements from Hinode EIS have provided new insights into the properties of SADs, validating earlier results and hypotheses, whilst providing new observables that future theory must incorporate. Our key findings from this dataset are:

\begin{enumerate}
    \item In the sit-and-stare data, SADs are observed as spatially and temporally-resolved drops in intensity, aligned with strong Doppler downflows (especially clear for the most prominent intensity drops). This suggests that the dark SADs themselves are downflowing, and not just low-density voids behind contracting loops, as suggested by \citet{Savage2012}.
    
    \item We see a pattern of stronger LOS downflows in the north of our region of interest, than in the south. This is suggestive of SADs diverging above the flare looptop. This is a similar picture to that presented in the cartoon in Figure 1 of \citet{Polito2018}, where Fe XXI blue and red shift patterns above a flare looptop suggest a scenario of reconnection outflows diverging beneath a high-altitude termination shock in a solar flare. The 200/250 km/s red/blue-shift velocities examined in \citet{Polito2018} are an order of magnitude higher than those presented in our study, but were measured closer to disk center, avoiding the projection effects present for our observations on the limb. Our observations could reveal a similar story, providing evidence for SADs diverging beneath a high altitude termination shock, as predicted by \citet{Shen2022}. We outline the suspected viewing geometry that allows us to view this SAD divergence, in the cartoon in Figure \ref{fig:SAD_cartoon}. This interpretation is \textit{not} in contradiction with the interpretation that SADs are contracting loops outflowing from the reconnection site (see references in the introduction), but can explain the disparity between SAD velocities and the local Alfv\'en speed in the lower solar corona, as the reconnection outflows are slowed down by the termination shock region.
    
    \item Statistically, the SADs exhibit higher non-thermal velocities than seen in the surrounding flare fan. The highest-contrast SADs show a time-lag between intensity drop and non-thermal velocity enhancement, which persists beyond the SADs passage across the slit location. This is in support of previous works from \citet{Shen2022} and \citet{Xie2025} that suggest SADs produce, and are comprised of, turbulent plasma. 
    
    \item Utilizing measurements from the temperature sensitive Fe XXIV 255.11 \AA\ and Fe XXIII 263.41 \AA\ pair, we find SAD temperatures in the range of 12.2-13.4 MK, close to that of the surrounding flare fan. These are in line with earlier temperature measurements from DEM analysis \citep{Hanneman2014,Tan2022}.
    
    \item We construct the 3D velocity field of four SADs, to find velocities primarily constrained in the POS, with magnitudes similar to those found in previous imaging studies \citep{McKenzie1999,McKenzie2009}. We find no evidence of more rapid velocities, such as the 800-1000 km/s downflows observed in \citet{Innes2003b}. Because our observations are limited to a sit-and-stare sequence, we cannot rule out that much faster velocities may exist at higher altitudes in the flare, such as those that may be found closer to the reconnection site, perhaps even above the termination shock region that rapidly decelerates the plasma.

    \item As per item 1, the dark SADs correspond spatially to red-shift structures. However, equivalent Doppler features, near identical in appearance to those produced by SADs, also exist elsewhere in the raster - at times and locations with no corresponding intensity drops in either the EIS intensity raster or AIA pseudo-raster. This suggests the existence of `stealth SADs' - downflowing SAD structures that produce no observable drop in intensity.
\end{enumerate}

Items 2--5 in the enumerated list above provide new evidence in support of pre-existing hypotheses of the nature of SADs. Items 1 and 6, however, are harder to ratify with previous literature on the subject. In the work of \citet{Savage2012}, the authors utilize high resolution AIA imaging to detect thin contracting loops at the front of downflowing SAD structures. The authors interpreted the results to suggest that the SADs themselves are not contracting flux tubes, but rather low-density voids in the wake of much finer contracting loop structures (called SADLs). However, in our observations, the strongest intensity drops coincide with strong red-shift pixels. Both intensity and red-shift features are spatially (on the vertical axis) and temporally (on the horizontal axis) resolved, revealing that the SADs themselves are downflowing, and not a static low-density void behind a thin downflowing loop/SADL. At first, this seems contradictory to the convincing evidence presented by \citet{Savage2012}. Possible explanations for this discrepancy could be that, in our event, the contracting loops at the forefront of each SAD are too thin to resolve spatially with either the EIS or AIA observations. This is possible, as \citet{Savage2012} state that such loops are only visible in AIA for the largest SADs, and not smaller features. None of the SADs present in our event are considerably large, so this resolution issue is a possible explanation. A second reason for this disparity could be that, unlike the fan and SAD emission, the contracting SADL loops are not emitting at the temperatures needed to produce Fe XXIV emission. We note, however, that we see no other cotemporal/cospatial narrow SAD-like dimmings or brightenings in any of the major lines in the EIS raster. Although these options allow our results to remain compatible with the interpretation of SADs from \citet{Savage2012} (that SADs are low-density regions behind contracting loops), it does not change the fact that we find clear evidence that the SADs we observe are red-shifted, with total velocities (considering both POS and LOS components) similar to that observed in previous observations of SADs. If SADs are low-density voids behind narrow contracting loops, our work suggests that they are not static, but still flowing towards the flare arcade. To our knowledge, this behavior is not yet explained by the model of SADs presented by \citet{Savage2012} and \citet{Scott2016}. 
Alternatively, the spatial alignment between intensity drops and Doppler downflows is more in agreement with earlier interpretations of SADs \citep[e.g. from][]{McKenzie1999,Savage2011}, in which the entire SAD void is a contracting low-density flux tube \citep[with a much larger cross-sectional area than the SADLs analyzed by][]{Savage2012}. However, this earlier interpretation now seems less likely, given the aforementioned work from \citet{Savage2011}.

The last puzzling result (item 6 in the enumerated list), is the presence of SAD-like velocity features with no corresponding drop in intensity. These `stealth SADs' have, to our knowledge, not been reported before. Given that SADs have been found to have a temperature similar to that in the surrounding fan \citep[][and from our analysis in this paper]{Hanneman2014}, the lower-intensity appearance of regular SADs in EUV must be due to a lower density than their surrounding plasma. However, with the existence of `stealth SADs', some SAD-like loop contractions must also occur without creating the same density drop that allows the visibility of regular SADs. This is another puzzle that must be explained by future modeling of these features.

\section{Conclusion}

Collecting spectral observations of SADs has proven to be very difficult over the past two decades and beyond, in part due to their small area, dependency on viewing angle, and difficulties in catching off-limb flares with limited FOV spectral instruments. The observations presented in this paper are therefore a special dataset, allowing the validation of previous interpretations of SADs, whilst also revealing new behaviors that must be incorporated into future theory and modeling. These findings include the spatial alignment between intensity-drops and Doppler velocity-enhancements in SADs, a divergence of SADs above the flare looptop, the turbulent nature of SADs, verifying previous SAD imaging temperature measurements, and the existence of `stealth SADs'.
Despite their elusive nature, the likely connection between SADs, flare reconnection site, and possible high-altitude termination shock, means they continue to offer potential insights into the fundamental energy release and particle acceleration processes in solar flares.

\begin{acknowledgments}
RJF thanks support from the Brinson Prize Fellowship. M.D.K. acknowledges support by NASA ECIP award 80NSSC19K0910 and NSF CAREER award SPVKK1RC2MZ3. KKR is supported by NSF grant AGS-2334929. Hinode is a Japanese mission developed and launched by ISAS/JAXA, with NAOJ as domestic partner and NASA and UKSA as international partners. It is operated by these agencies in co-operation with ESA and NSC (Norway).
\end{acknowledgments}

\bibliography{bibliography}

\begin{thebibliography}{}
\expandafter\ifx\csname natexlab\endcsname\relax\def\natexlab#1{#1}\fi
\providecommand{\url}[1]{\href{#1}{#1}}
\providecommand{\dodoi}[1]{doi:~\href{http://doi.org/#1}{\nolinkurl{#1}}}
\providecommand{\doeprint}[1]{\href{http://ascl.net/#1}{\nolinkurl{http://ascl.net/#1}}}
\providecommand{\doarXiv}[1]{\href{https://arxiv.org/abs/#1}{\nolinkurl{https://arxiv.org/abs/#1}}}

\bibitem[{M. {Br{\"o}se} {et~al.}(2022){Br{\"o}se}, {Warmuth}, {Sakao}, \& {Su}}]{Brose2022}
{Br{\"o}se}, M., {Warmuth}, A., {Sakao}, T., \& {Su}, Y. 2022, \bibinfo{title}{{Temperature and differential emission measure evolution of a limb flare on 13 January 2015},} \aap, 663, A18, \dodoi{10.1051/0004-6361/202141868}

\bibitem[{H. {Carmichael}(1964){Carmichael}}]{Carmichael1964}
{Carmichael}, H. 1964, \bibinfo{title}{{A Process for Flares},} NASA Special Publication, 50, 451

\bibitem[{M. {C{\'e}cere} {et~al.}(2012){C{\'e}cere}, {Schneiter}, {Costa}, {Elaskar}, \& {Maglione}}]{Cecere2012}
{C{\'e}cere}, M., {Schneiter}, M., {Costa}, A., {Elaskar}, S., \& {Maglione}, S. 2012, \bibinfo{title}{{Simulation of Descending Multiple Supra-arcade Reconnection Outflows in Solar Flares},} \apj, 759, 79, \dodoi{10.1088/0004-637X/759/2/79}

\bibitem[{B. {Chen} {et~al.}(2015){Chen}, {Bastian}, {Shen}, {Gary}, {Krucker}, \& {Glesener}}]{Chen2015}
{Chen}, B., {Bastian}, T.~S., {Shen}, C., {et~al.} 2015, \bibinfo{title}{{Particle acceleration by a solar flare termination shock},} Science, 350, 1238, \dodoi{10.1126/science.aac8467}

\bibitem[{B. {Chen} {et~al.}(2019){Chen}, {Shen}, {Reeves}, {Guo}, \& {Yu}}]{Chen2019}
{Chen}, B., {Shen}, C., {Reeves}, K.~K., {Guo}, F., \& {Yu}, S. 2019, \bibinfo{title}{{Radio Spectroscopic Imaging of a Solar Flare Termination Shock: Split-band Feature as Evidence for Shock Compression},} \apj, 884, 63, \dodoi{10.3847/1538-4357/ab3c58}

\bibitem[{J.~L. {Culhane} {et~al.}(2007){Culhane}, {Harra}, {James}, {Al-Janabi}, {Bradley}, {Chaudry}, {Rees}, {Tandy}, {Thomas}, {Whillock}, {Winter}, {Doschek}, {Korendyke}, {Brown}, {Myers}, {Mariska}, {Seely}, {Lang}, {Kent}, {Shaughnessy}, {Young}, {Simnett}, {Castelli}, {Mahmoud}, {Mapson-Menard}, {Probyn}, {Thomas}, {Davila}, {Dere}, {Windt}, {Shea}, {Hagood}, {Moye}, {Hara}, {Watanabe}, {Matsuzaki}, {Kosugi}, {Hansteen}, \& {Wikstol}}]{Culhane2007}
{Culhane}, J.~L., {Harra}, L.~K., {James}, A.~M., {et~al.} 2007, \bibinfo{title}{{The EUV Imaging Spectrometer for Hinode},} \solphys, 243, 19, \dodoi{10.1007/s01007-007-0293-1}

\bibitem[{K.~P. {Dere} {et~al.}(1997){Dere}, {Landi}, {Mason}, {Monsignori Fossi}, \& {Young}}]{Dere1997}
{Dere}, K.~P., {Landi}, E., {Mason}, H.~E., {Monsignori Fossi}, B.~C., \& {Young}, P.~R. 1997, \bibinfo{title}{{CHIANTI - an atomic database for emission lines},} \aaps, 125, 149, \dodoi{10.1051/aas:1997368}

\bibitem[{G.~A. {Doschek} {et~al.}(2014){Doschek}, {McKenzie}, \& {Warren}}]{Doschek2014}
{Doschek}, G.~A., {McKenzie}, D.~E., \& {Warren}, H.~P. 2014, \bibinfo{title}{{Plasma Dynamics Above Solar Flare Soft X-Ray Loop Tops},} \apj, 788, 26, \dodoi{10.1088/0004-637X/788/1/26}

\bibitem[{R.~P. {Dufresne} {et~al.}(2024){Dufresne}, {Del Zanna}, {Young}, {Dere}, {Deliporanidou}, {Barnes}, \& {Landi}}]{Dufresne2024}
{Dufresne}, R.~P., {Del Zanna}, G., {Young}, P.~R., {et~al.} 2024, \bibinfo{title}{{CHIANTI{\textemdash}An Atomic Database for Emission Lines{\textemdash}Paper. XVIII. Version 11, Advanced Ionization Equilibrium Models: Density and Charge Transfer Effects},} \apj, 974, 71, \dodoi{10.3847/1538-4357/ad6765}

\bibitem[{R.~J. {French} {et~al.}(2024{\natexlab{a}}){French}, {Hayes}, {Kazachenko}, {Reeves}, {Shen}, \& {L{\"o}rin{\v{c}}{\'\i}k}}]{French2024b}
{French}, R.~J., {Hayes}, L.~A., {Kazachenko}, M.~D., {et~al.} 2024{\natexlab{a}}, \bibinfo{title}{{X-Ray and Spectral Ultraviolet Observations of Periodic Pulsations in a Solar Flare Fan/Looptop},} \apj, 977, 207, \dodoi{10.3847/1538-4357/ad8ed1}

\bibitem[{R.~J. {French} {et~al.}(2020){French}, {Matthews}, {van Driel-Gesztelyi}, {Long}, \& {Judge}}]{French2020}
{French}, R.~J., {Matthews}, S.~A., {van Driel-Gesztelyi}, L., {Long}, D.~M., \& {Judge}, P.~G. 2020, \bibinfo{title}{{Dynamics of Late-stage Reconnection in the 2017 September 10 Solar Flare},} \apj, 900, 192, \dodoi{10.3847/1538-4357/aba94b}

\bibitem[{R.~J. {French} {et~al.}(2024{\natexlab{b}}){French}, {Yu}, {Chen}, {Shen}, \& {Matthews}}]{French2024a}
{French}, R.~J., {Yu}, S., {Chen}, B., {Shen}, C., \& {Matthews}, S.~A. 2024{\natexlab{b}}, \bibinfo{title}{{Doppler signature of a possible termination shock in an off-limb solar flare},} \mnras, 528, 6836, \dodoi{10.1093/mnras/stae430}

\bibitem[{B.~N. {Handy} {et~al.}(1999){Handy}, {Acton}, {Kankelborg}, {Wolfson}, {Akin}, {Bruner}, {Caravalho}, {Catura}, {Chevalier}, {Duncan}, {Edwards}, {Feinstein}, {Freeland}, {Friedlaender}, {Hoffmann}, {Hurlburt}, {Jurcevich}, {Katz}, {Kelly}, {Lemen}, {Levay}, {Lindgren}, {Mathur}, {Meyer}, {Morrison}, {Morrison}, {Nightingale}, {Pope}, {Rehse}, {Schrijver}, {Shine}, {Shing}, {Strong}, {Tarbell}, {Title}, {Torgerson}, {Golub}, {Bookbinder}, {Caldwell}, {Cheimets}, {Davis}, {Deluca}, {McMullen}, {Warren}, {Amato}, {Fisher}, {Maldonado}, \& {Parkinson}}]{Handy1999}
{Handy}, B.~N., {Acton}, L.~W., {Kankelborg}, C.~C., {et~al.} 1999, \bibinfo{title}{{The transition region and coronal explorer},} \solphys, 187, 229, \dodoi{10.1023/A:1005166902804}

\bibitem[{W.~J. {Hanneman} \& K.~K. {Reeves}(2014){Hanneman} \& {Reeves}}]{Hanneman2014}
{Hanneman}, W.~J., \& {Reeves}, K.~K. 2014, \bibinfo{title}{{Thermal Structure of Current Sheets and Supra-arcade Downflows in the Solar Corona},} \apj, 786, 95, \dodoi{10.1088/0004-637X/786/2/95}

\bibitem[{T. {Hirayama}(1974){Hirayama}}]{Hirayama1974}
{Hirayama}, T. 1974, \bibinfo{title}{{Theoretical Model of Flares and Prominences. I: Evaporating Flare Model},} \solphys, 34, 323, \dodoi{10.1007/BF00153671}

\bibitem[{D.~E. {Innes} {et~al.}(2003{\natexlab{a}}){Innes}, {McKenzie}, \& {Wang}}]{Innes2003a}
{Innes}, D.~E., {McKenzie}, D.~E., \& {Wang}, T. 2003{\natexlab{a}}, \bibinfo{title}{{SUMER spectral observations of post-flare supra-arcade inflows},} \solphys, 217, 247, \dodoi{10.1023/B:SOLA.0000006899.12788.22}

\bibitem[{D.~E. {Innes} {et~al.}(2003{\natexlab{b}}){Innes}, {McKenzie}, \& {Wang}}]{Innes2003b}
{Innes}, D.~E., {McKenzie}, D.~E., \& {Wang}, T. 2003{\natexlab{b}}, \bibinfo{title}{{Observations of 1000 km s$^{{\ensuremath{-}}1}$ Doppler shifts in {}10$^{7}$ K solar flare supra-arcade},} \solphys, 217, 267, \dodoi{10.1023/B:SOLA.0000006874.31799.bc}

\bibitem[{M. {Janvier} {et~al.}(2023){Janvier}, {Mzerguat}, {Young}, {Buchlin}, {Manou}, {Pelouze}, {Long}, {Green}, {Warmuth}, {Schuller}, {D{\'e}moulin}, {Calchetti}, {Kahil}, {Bellot Rubio}, {Parenti}, {Baccar}, {Barczynski}, {Harra}, {Hayes}, {Thompson}, {M{\"u}ller}, {Baker}, {Yardley}, {Berghmans}, {Verbeeck}, {Smith}, {Peter}, {Aznar Cuadrado}, {Musset}, {Brooks}, {Rodr{\'\i}guez}, {Auch{\`e}re}, {Carlsson}, {Fludra}, {Hassler}, {Williams}, {Caldwell}, {Fredvik}, {Giunta}, {Grundy}, {Guest}, {Kraaikamp}, {Leeks}, {Plowman}, {Schmutz}, {Sch{\"u}hle}, {Sidher}, {Teriaca}, {Solanki}, {del Toro Iniesta}, {Woch}, {Gandorfer}, {Hirzberger}, {Orozco Su{\'a}rez}, {Appourchaux}, {Valori}, {Sinjan}, {Albert}, \& {Volkmer}}]{Janvier2023}
{Janvier}, M., {Mzerguat}, S., {Young}, P.~R., {et~al.} 2023, \bibinfo{title}{{A multiple spacecraft detection of the 2 April 2022 M-class flare and filament eruption during the first close Solar Orbiter perihelion},} \aap, 677, A130, \dodoi{10.1051/0004-6361/202346321}

\bibitem[{T. {Kawate} {et~al.}(2016){Kawate}, {Keenan}, \& {Jess}}]{Kawate2016}
{Kawate}, T., {Keenan}, F.~P., \& {Jess}, D.~B. 2016, \bibinfo{title}{{Departure of High-temperature Iron Lines from the Equilibrium State in Flaring Solar Plasmas},} \apj, 826, 3, \dodoi{10.3847/0004-637X/826/1/3}

\bibitem[{J.~I. {Khan} {et~al.}(2007){Khan}, {Bain}, \& {Fletcher}}]{Khan2007}
{Khan}, J.~I., {Bain}, H.~M., \& {Fletcher}, L. 2007, \bibinfo{title}{{The relative timing of supra-arcade downflows in solar flares},} \aap, 475, 333, \dodoi{10.1051/0004-6361:20077894}

\bibitem[{R.~A. {Kopp} \& G.~W. {Pneuman}(1976){Kopp} \& {Pneuman}}]{Kopp1976}
{Kopp}, R.~A., \& {Pneuman}, G.~W. 1976, \bibinfo{title}{{Magnetic reconnection in the corona and the loop prominence phenomenon.},} \solphys, 50, 85, \dodoi{10.1007/BF00206193}

\bibitem[{S. {Krucker} {et~al.}(2020){Krucker}, {Hurford}, {Grimm}, {K{\"o}gl}, {Gr{\"o}belbauer}, {Etesi}, {Casadei}, {Csillaghy}, {Benz}, {Arnold}, {Molendini}, {Orleanski}, {Schori}, {Xiao}, {Kuhar}, {Hochmuth}, {Felix}, {Schramka}, {Marcin}, {Kobler}, {Iseli}, {Dreier}, {Wiehl}, {Kleint}, {Battaglia}, {Lastufka}, {Sathiapal}, {Lapadula}, {Bednarzik}, {Birrer}, {Stutz}, {Wild}, {Marone}, {Skup}, {Cichocki}, {Ber}, {Rutkowski}, {Bujwan}, {Juchnikowski}, {Winkler}, {Darmetko}, {Michalska}, {Seweryn}, {Bia{\l}ek}, {Osica}, {Sylwester}, {Kowalinski}, {{\'S}cis{\l}owski}, {Siarkowski}, {St{\k{e}}{\'s}licki}, {Mrozek}, {Podg{\'o}rski}, {Meuris}, {Limousin}, {Gevin}, {Le Mer}, {Brun}, {Strugarek}, {Vilmer}, {Musset}, {Maksimovi{\'c}}, {F{\'a}rn{\'\i}k}, {Koz{\'a}{\v{c}}ek}, {Ka{\v{s}}parov{\'a}}, {Mann}, {{\"O}nel}, {Warmuth}, {Rendtel}, {Anderson}, {Bauer}, {Dionies}, {Paschke}, {Pl{\"u}schke}, {Woche}, {Schuller}, {Veronig}, {Dickson}, {Gallagher}, {Maloney}, {Bloomfield}, {Piana}, {Massone}, {Benvenuto},
  {Massa}, {Schwartz}, {Dennis}, {van Beek}, {Rodr{\'\i}guez-Pacheco}, \& {Lin}}]{Krucker2020}
{Krucker}, S., {Hurford}, G.~J., {Grimm}, O., {et~al.} 2020, \bibinfo{title}{{The Spectrometer/Telescope for Imaging X-rays (STIX)},} \aap, 642, A15, \dodoi{10.1051/0004-6361/201937362}

\bibitem[{E. {Kupriyanova} {et~al.}(2020){Kupriyanova}, {Kolotkov}, {Nakariakov}, \& {Kaufman}}]{Kupriyanova2020}
{Kupriyanova}, E., {Kolotkov}, D., {Nakariakov}, V., \& {Kaufman}, A. 2020, \bibinfo{title}{{Quasi-Periodic Pulsations in Solar and Stellar Flares. Review},} Solar-Terrestrial Physics, 6, 3, \dodoi{10.12737/stp-61202001}

\bibitem[{J.~R. {Lemen} {et~al.}(2012){Lemen}, {Title}, {Akin}, {Boerner}, {Chou}, {Drake}, {Duncan}, {Edwards}, {Friedlaender}, {Heyman}, {Hurlburt}, {Katz}, {Kushner}, {Levay}, {Lindgren}, {Mathur}, {McFeaters}, {Mitchell}, {Rehse}, {Schrijver}, {Springer}, {Stern}, {Tarbell}, {Wuelser}, {Wolfson}, {Yanari}, {Bookbinder}, {Cheimets}, {Caldwell}, {Deluca}, {Gates}, {Golub}, {Park}, {Podgorski}, {Bush}, {Scherrer}, {Gummin}, {Smith}, {Auker}, {Jerram}, {Pool}, {Soufli}, {Windt}, {Beardsley}, {Clapp}, {Lang}, \& {Waltham}}]{Lemen2012}
{Lemen}, J.~R., {Title}, A.~M., {Akin}, D.~J., {et~al.} 2012, \bibinfo{title}{{The Atmospheric Imaging Assembly (AIA) on the Solar Dynamics Observatory (SDO)},} \solphys, 275, 17, \dodoi{10.1007/s11207-011-9776-8}

\bibitem[{Z.~F. {Li} {et~al.}(2021){Li}, {Cheng}, {Ding}, {Reeves}, {Kittrell}, {Weber}, \& {McKenzie}}]{Li2021}
{Li}, Z.~F., {Cheng}, X., {Ding}, M.~D., {et~al.} 2021, \bibinfo{title}{{Thermodynamic Evolution of Solar Flare Supra-arcade Downflows},} \apj, 915, 124, \dodoi{10.3847/1538-4357/ac043e}

\bibitem[{D. {Longcope} {et~al.}(2018){Longcope}, {Unverferth}, {Klein}, {McCarthy}, \& {Priest}}]{Longcope2018}
{Longcope}, D., {Unverferth}, J., {Klein}, C., {McCarthy}, M., \& {Priest}, E. 2018, \bibinfo{title}{{Evidence for Downflows in the Narrow Plasma Sheet of 2017 September 10 and Their Significance for Flare Reconnection},} \apj, 868, 148, \dodoi{10.3847/1538-4357/aaeac4}

\bibitem[{L.~S. {Maglione} {et~al.}(2011){Maglione}, {Schneiter}, {Costa}, \& {Elaskar}}]{Maglione2011}
{Maglione}, L.~S., {Schneiter}, E.~M., {Costa}, A., \& {Elaskar}, S. 2011, \bibinfo{title}{{Simulation of dark lanes in post-flare supra-arcades. III. A 2D simulation},} \aap, 527, L5, \dodoi{10.1051/0004-6361/201015934}

\bibitem[{J.~T. {Mariska} \& K. {Muglach}(2010){Mariska} \& {Muglach}}]{Mariska2010}
{Mariska}, J.~T., \& {Muglach}, K. 2010, \bibinfo{title}{{Doppler-shift, Intensity, and Density Oscillations Observed with the Extreme Ultraviolet Imaging Spectrometer on Hinode},} \apj, 713, 573, \dodoi{10.1088/0004-637X/713/1/573}

\bibitem[{D.~E. {McKenzie}(2000){McKenzie}}]{McKenzie2000}
{McKenzie}, D.~E. 2000, \bibinfo{title}{{Supra-arcade Downflows in Long-Duration Solar Flare Events},} \solphys, 195, 381, \dodoi{10.1023/A:1005220604894}

\bibitem[{D.~E. {McKenzie} \& H.~S. {Hudson}(1999){McKenzie} \& {Hudson}}]{McKenzie1999}
{McKenzie}, D.~E., \& {Hudson}, H.~S. 1999, \bibinfo{title}{{X-Ray Observations of Motions and Structure above a Solar Flare Arcade},} \apjl, 519, L93, \dodoi{10.1086/312110}

\bibitem[{D.~E. {McKenzie} \& S.~L. {Savage}(2009){McKenzie} \& {Savage}}]{McKenzie2009}
{McKenzie}, D.~E., \& {Savage}, S.~L. 2009, \bibinfo{title}{{Quantitative Examination of Supra-arcade Downflows in Eruptive Solar Flares},} \apj, 697, 1569, \dodoi{10.1088/0004-637X/697/2/1569}

\bibitem[{A.~R. {Miles}(2009){Miles}}]{Miles2009}
{Miles}, A.~R. 2009, \bibinfo{title}{{THE Blast-Wave-Driven Instability as a Vehicle for Understanding Supernova Explosion Structure},} \apj, 696, 498, \dodoi{10.1088/0004-637X/696/1/498}

\bibitem[{D. {M{\"u}ller} {et~al.}(2020){M{\"u}ller}, {St. Cyr}, {Zouganelis}, {Gilbert}, {Marsden}, {Nieves-Chinchilla}, {Antonucci}, {Auch{\`e}re}, {Berghmans}, {Horbury}, {Howard}, {Krucker}, {Maksimovic}, {Owen}, {Rochus}, {Rodriguez-Pacheco}, {Romoli}, {Solanki}, {Bruno}, {Carlsson}, {Fludra}, {Harra}, {Hassler}, {Livi}, {Louarn}, {Peter}, {Sch{\"u}hle}, {Teriaca}, {del Toro Iniesta}, {Wimmer-Schweingruber}, {Marsch}, {Velli}, {De Groof}, {Walsh}, \& {Williams}}]{Muller2020}
{M{\"u}ller}, D., {St. Cyr}, O.~C., {Zouganelis}, I., {et~al.} 2020, \bibinfo{title}{{The Solar Orbiter mission. Science overview},} \aap, 642, A1, \dodoi{10.1051/0004-6361/202038467}

\bibitem[{V.~M. {Nakariakov} \& V.~F. {Melnikov}(2009){Nakariakov} \& {Melnikov}}]{nakariakov2009}
{Nakariakov}, V.~M., \& {Melnikov}, V.~F. 2009, \bibinfo{title}{{Quasi-Periodic Pulsations in Solar Flares},} \ssr, 149, 119, \dodoi{10.1007/s11214-009-9536-3}

\bibitem[{V. {Polito} {et~al.}(2018){Polito}, {Galan}, {Reeves}, \& {Musset}}]{Polito2018}
{Polito}, V., {Galan}, G., {Reeves}, K.~K., \& {Musset}, S. 2018, \bibinfo{title}{{Possible Signatures of a Termination Shock in the 2014 March 29 X-class Flare Observed by IRIS},} \apj, 865, 161, \dodoi{10.3847/1538-4357/aadada}

\bibitem[{K.~K. {Reeves} {et~al.}(2017){Reeves}, {Freed}, {McKenzie}, \& {Savage}}]{Reeves2017}
{Reeves}, K.~K., {Freed}, M.~S., {McKenzie}, D.~E., \& {Savage}, S.~L. 2017, \bibinfo{title}{{An Exploration of Heating Mechanisms in a Supra-arcade Plasma Sheet Formed after a Coronal Mass Ejection},} \apj, 836, 55, \dodoi{10.3847/1538-4357/836/1/55}

\bibitem[{K.~K. {Reeves} {et~al.}(2020){Reeves}, {Polito}, {Chen}, {Galan}, {Yu}, {Liu}, \& {Li}}]{Reeves2020}
{Reeves}, K.~K., {Polito}, V., {Chen}, B., {et~al.} 2020, \bibinfo{title}{{Hot Plasma Flows and Oscillations in the Loop-top Region During the 2017 September 10 X8.2 Solar Flare},} \apj, 905, 165, \dodoi{10.3847/1538-4357/abc4e0}

\bibitem[{T. {Samanta} {et~al.}(2021){Samanta}, {Tian}, {Chen}, {Reeves}, {Cheung}, {Vourlidas}, \& {Banerjee}}]{Samanta2021}
{Samanta}, T., {Tian}, H., {Chen}, B., {et~al.} 2021, \bibinfo{title}{{Plasma heating induced by tadpole-like downflows in the flaring solar corona},} The Innovation, 2, 100083, \dodoi{10.1016/j.xinn.2021.100083}

\bibitem[{S.~L. {Savage} \& D.~E. {McKenzie}(2011){Savage} \& {McKenzie}}]{Savage2011}
{Savage}, S.~L., \& {McKenzie}, D.~E. 2011, \bibinfo{title}{{Quantitative Examination of a Large Sample of Supra-arcade Downflows in Eruptive Solar Flares},} \apj, 730, 98, \dodoi{10.1088/0004-637X/730/2/98}

\bibitem[{S.~L. {Savage} {et~al.}(2012){Savage}, {McKenzie}, \& {Reeves}}]{Savage2012}
{Savage}, S.~L., {McKenzie}, D.~E., \& {Reeves}, K.~K. 2012, \bibinfo{title}{{Re-interpretation of Supra-arcade Downflows in Solar Flares},} \apjl, 747, L40, \dodoi{10.1088/2041-8205/747/2/L40}

\bibitem[{R.~B. {Scott} {et~al.}(2016){Scott}, {McKenzie}, \& {Longcope}}]{Scott2016}
{Scott}, R.~B., {McKenzie}, D.~E., \& {Longcope}, D.~W. 2016, \bibinfo{title}{{Inferring the Magnetohydrodynamic Structure of Solar Flare Supra-Arcade Plasmas from a Data-assimilated Field Transport Model},} \apj, 819, 56, \dodoi{10.3847/0004-637X/819/1/56}

\bibitem[{C. {Shen} {et~al.}(2022){Shen}, {Chen}, {Reeves}, {Yu}, {Polito}, \& {Xie}}]{Shen2022}
{Shen}, C., {Chen}, B., {Reeves}, K.~K., {et~al.} 2022, \bibinfo{title}{{The origin of underdense plasma downflows associated with magnetic reconnection in solar flares},} Nature Astronomy, 6, 317, \dodoi{10.1038/s41550-021-01570-2}

\bibitem[{C. {Shen} {et~al.}(2011){Shen}, {Lin}, \& {Murphy}}]{Shen2011}
{Shen}, C., {Lin}, J., \& {Murphy}, N.~A. 2011, \bibinfo{title}{{Numerical Experiments on Fine Structure within Reconnecting Current Sheets in Solar Flares},} \apj, 737, 14, \dodoi{10.1088/0004-637X/737/1/14}

\bibitem[{C. {Shen} {et~al.}(2023){Shen}, {Polito}, {Reeves}, {Chen}, {Yu}, \& {Xie}}]{Shen2023}
{Shen}, C., {Polito}, V., {Reeves}, K.~K., {et~al.} 2023, \bibinfo{title}{{Non-thermal Broadening of IRIS Fe XXI Line Caused by Turbulent Plasma Flows in the Magnetic Reconnection Region During Solar Eruptions},} Frontiers in Astronomy and Space Sciences, 10, 19, \dodoi{10.3389/fspas.2023.1096133}

\bibitem[{P.~A. {Sturrock}(1968){Sturrock}}]{Sturrock1968}
{Sturrock}, P.~A. 1968, in Structure and Development of Solar Active Regions, ed. K.~O. {Kiepenheuer}, Vol.~35, 471

\bibitem[{G. {Tan} {et~al.}(2022){Tan}, {Hou}, \& {Tian}}]{Tan2022}
{Tan}, G., {Hou}, Y., \& {Tian}, H. 2022, \bibinfo{title}{{Statistical investigation of the kinematic and thermal properties of supra-arcade downflows observed during a solar flare},} \mnras, 516, 3120, \dodoi{10.1093/mnras/stac2470}

\bibitem[{S. {Tsuneta} {et~al.}(1991){Tsuneta}, {Acton}, {Bruner}, {Lemen}, {Brown}, {Caravalho}, {Catura}, {Freeland}, {Jurcevich}, {Morrison}, {Ogawara}, {Hirayama}, \& {Owens}}]{Tsuneta1991}
{Tsuneta}, S., {Acton}, L., {Bruner}, M., {et~al.} 1991, \bibinfo{title}{{The Soft X-ray Telescope for the SOLAR-A mission},} \solphys, 136, 37, \dodoi{10.1007/BF00151694}

\bibitem[{T. {Van Doorsselaere} {et~al.}(2016){Van Doorsselaere}, {Kupriyanova}, \& {Yuan}}]{vandoorsselare2016}
{Van Doorsselaere}, T., {Kupriyanova}, E.~G., \& {Yuan}, D. 2016, \bibinfo{title}{{Quasi-periodic Pulsations in Solar and Stellar Flares: An Overview of Recent Results (Invited Review)},} \solphys, 291, 3143, \dodoi{10.1007/s11207-016-0977-z}

\bibitem[{H.~P. {Warren} {et~al.}(2018){Warren}, {Brooks}, {Ugarte-Urra}, {Reep}, {Crump}, \& {Doschek}}]{Warren2018}
{Warren}, H.~P., {Brooks}, D.~H., {Ugarte-Urra}, I., {et~al.} 2018, \bibinfo{title}{{Spectroscopic Observations of Current Sheet Formation and Evolution},} \apj, 854, 122, \dodoi{10.3847/1538-4357/aaa9b8}

\bibitem[{H.~P. {Warren} {et~al.}(2011){Warren}, {O'Brien}, \& {Sheeley}}]{Warren2011}
{Warren}, H.~P., {O'Brien}, C.~M., \& {Sheeley}, Jr., N.~R. 2011, \bibinfo{title}{{Observations of Reconnecting Flare Loops with the Atmospheric Imaging Assembly},} \apj, 742, 92, \dodoi{10.1088/0004-637X/742/2/92}

\bibitem[{J.~S. {Warren} {et~al.}(2005){Warren}, {Hughes}, {Badenes}, {Ghavamian}, {McKee}, {Moffett}, {Plucinsky}, {Rakowski}, {Reynoso}, \& {Slane}}]{Warren2005}
{Warren}, J.~S., {Hughes}, J.~P., {Badenes}, C., {et~al.} 2005, \bibinfo{title}{{Cosmic-Ray Acceleration at the Forward Shock in Tycho's Supernova Remnant: Evidence from Chandra X-Ray Observations},} \apj, 634, 376, \dodoi{10.1086/496941}

\bibitem[{K. {Wilhelm} {et~al.}(1995){Wilhelm}, {Curdt}, {Marsch}, {Sch{\"u}hle}, {Lemaire}, {Gabriel}, {Vial}, {Grewing}, {Huber}, {Jordan}, {Poland}, {Thomas}, {K{\"u}hne}, {Timothy}, {Hassler}, \& {Siegmund}}]{Wilhelm1995}
{Wilhelm}, K., {Curdt}, W., {Marsch}, E., {et~al.} 1995, \bibinfo{title}{{SUMER - Solar Ultraviolet Measurements of Emitted Radiation},} \solphys, 162, 189, \dodoi{10.1007/BF00733430}

\bibitem[{X. {Xie} \& K.~K. {Reeves}(2023){Xie} \& {Reeves}}]{Xie2023}
{Xie}, X., \& {Reeves}, K.~K. 2023, \bibinfo{title}{{Heating Effects of Supra-arcade Downflows on Plasma above Solar Flare Arcades},} \apj, 942, 28, \dodoi{10.3847/1538-4357/ac9f47}

\bibitem[{X. {Xie} {et~al.}(2022){Xie}, {Reeves}, {Shen}, \& {Ingram}}]{Xie2022}
{Xie}, X., {Reeves}, K.~K., {Shen}, C., \& {Ingram}, J.~D. 2022, \bibinfo{title}{{Statistical Study of the Kinetic Features of Supra-arcade Downflows Detected from Multiple Solar Flares},} \apj, 933, 15, \dodoi{10.3847/1538-4357/ac695d}

\bibitem[{X. {Xie} {et~al.}(2025){Xie}, {Shen}, {Reeves}, {Chen}, {Li}, {Guo}, {Yu}, {Wei}, \& {Dong}}]{Xie2025}
{Xie}, X., {Shen}, C., {Reeves}, K., {et~al.} 2025, \bibinfo{title}{{Anisotropic Turbulent Flows Observed in Above the Loop-top Regions During Solar Flares},} arXiv e-prints, arXiv:2503.13827, \dodoi{10.48550/arXiv.2503.13827}

\bibitem[{J. {Xue} {et~al.}(2020){Xue}, {Su}, {Li}, \& {Zhao}}]{Xue2020}
{Xue}, J., {Su}, Y., {Li}, H., \& {Zhao}, X. 2020, \bibinfo{title}{{Thermodynamical Evolution of Supra-arcade Downflows},} \apj, 898, 88, \dodoi{10.3847/1538-4357/ab9a3d}

\bibitem[{P.~R. {Young} {et~al.}(2012){Young}, {O'Dwyer}, \& {Mason}}]{Young2012}
{Young}, P.~R., {O'Dwyer}, B., \& {Mason}, H.~E. 2012, \bibinfo{title}{{Velocity Measurements for a Solar Active Region Fan Loop from Hinode/EIS Observations},} \apj, 744, 14, \dodoi{10.1088/0004-637X/744/1/14}

\bibitem[{P.~R. {Young} {et~al.}(2007){Young}, {Del Zanna}, {Mason}, {Dere}, {Landi}, {Landini}, {Doschek}, {Brown}, {Culhane}, {Harra}, {Watanabe}, \& {Hara}}]{Young2007}
{Young}, P.~R., {Del Zanna}, G., {Mason}, H.~E., {et~al.} 2007, \bibinfo{title}{{EUV Emission Lines and Diagnostics Observed with Hinode/EIS},} \pasj, 59, S857, \dodoi{10.1093/pasj/59.sp3.S857}

\bibitem[{I.~V. {Zimovets} {et~al.}(2021){Zimovets}, {McLaughlin}, {Srivastava}, {Kolotkov}, {Kuznetsov}, {Kupriyanova}, {Cho}, {Inglis}, {Reale}, {Pascoe}, {Tian}, {Yuan}, {Li}, \& {Zhang}}]{Zimovets2021}
{Zimovets}, I.~V., {McLaughlin}, J.~A., {Srivastava}, A.~K., {et~al.} 2021, \bibinfo{title}{{Quasi-Periodic Pulsations in Solar and Stellar Flares: A Review of Underpinning Physical Mechanisms and Their Predicted Observational Signatures},} \ssr, 217, 66, \dodoi{10.1007/s11214-021-00840-9}

\end{thebibliography}

\bibliographystyle{aasjournal}



\end{document}